
\input phyzzx
\def\ls#1{_{\lower1.5pt\hbox{$\scriptstyle #1$}}}
\Pubnum{SCIPP 93/30}

\def\SCIPP{\centerline {\it Santa Cruz Institute for Particle Physics}
  \centerline{\it University of California, Santa Cruz, CA 95064}}
\overfullrule 0pt
\titlepage
\Pubnum{SCIPP 93/17}
\date{June, 1992}
\vskip3cm
\title{{Topics In String Phenomenology}
\foot{Invited Talk Presented at Conference STRINGS 93, University of
California, Berkeley
1993}}
\foot{Work supported in part by the U.S. Department of Energy.}
\author{Michael Dine
}
\SCIPP
\vskip.5cm
\vbox{
\centerline{\bf Abstract}

\parskip 0pt
\parindent 25pt
\overfullrule=0pt
\baselineskip=18pt
\tolerance 3500
\endpage
\pagenumber=1
\singlespace
\bigskip


We consider two questions in string ``phenomenology.''
 First,
are there any generic string predictions?
Second, are there any general lessons which string theory
suggests for thinking about low energy models, particularly
in the framework of supersymmetry?  Among the topics
we consider are the squark and slepton spectrum, flavor symmetries,
discrete symmetries including $CP$, and Peccei-Quinn symmetries.
We also note that in some cases, discrete symmetries can
be used to constrain the form of supersymmetry breaking.

}
\singlespace
\bigskip

\chapter{Introduction}
The title of this talk is rather presumptious.  In my view, we are far
from possessing any string phenomenology; we still have very little
idea how the observable world
might emerge from string theory.  This is not to minimize the
fact that many compactifications of string theory have been
discovered with desirable features:  standard model gauge groups,
three generations of quarks and leptons, low energy supersymmetry,
intricate patterns of discrete symmetry and more.  Yet in detail, it
is probably safe to say that all of these models have serious flaws.
Moreover, we don't presently have a clue as
to how or why one of these models might be picked out over
another or why the cosmological constant should vanish after
supersymmetry breaking.

As a result, my goals today will be far more modest.  I won't review
the many interesting efforts to develop a detailed string
phenomenology.  Rather I will
\pointbegin
Look for features of string compactifications which might be generic,
leading to qualitative or quantitative predictions
\point
Seek insights from string theory into more conventional model
building.

\noindent
Virtually all of what I have to say will be in the framework of
models with low energy supersymmetry.
\REF\vadim{V.S. Kaplunovsky and J. Louis,
CERN-TH. 6809/93 UTTG-05-93.}
\REF\dkl{M. Dine, A. Kagan and R. Leigh, SCIPP-93-05.}
\REF\nirseibergb{Y. Nir and N. Seiberg, Phys. Lett. {\bf B309} (1993) 307.}
\REF\seibergrecent{N. Seiberg and P. Pouliot, RU-93-39.}
In the first category,
I will consider the minimal supersymmetric standard model.
This model has become a paradigm for a low energy supersymmetric
world.  Yet it rests on a set of strong assumptions about the
underlying microscopic theory.  Generically, these assumptions are
not true in string theory.  Recently, however,
Kaplunovsky and Louis have noted that
there is one scenario for string dynamics
in which they {\it are} true.\refmark{\vadim}
In this scenario, one
obtains a two (or three) parameter description of the low
energy world, and predicts a significant degree of degeneracy
among squarks and sleptons (necessary to suppress rare processes).
We will also describe alternative,
string-inspired scenarios for obtaining
squark degeneracy based on flavor symmetries.\refmark{\dkl,
\nirseibergb,\seibergrecent}

\REF\banksdixon{T. Banks, L. Dixon, D. Friedan and E. Martinec, Nucl.
Phys. {\bf B299} (1988) 613.}
\REF\nirseiberg{M. Leurer, Y. Nir and N. Seiberg, Phys. Lett. {\bf B398}
(1993)319.}
\REF\dilatonduality{A. Font, L. Ibanez, D. Lust
and F. Quevedo, Phys. Lett. {\bf B249} (1990) 35;
S.J. Rey, Phys. Re. {\bf D43} (1991) 526; A. Sen,
preprint TIFR-Th-92-41;
J. Schwarz, CALT-68-1815; J.E. Cohn and V. Periwal,
IASSNS-HEP-93-24.}
\REF\shenker{S.H. Shenker, talk given at the
Cargese Workshop on Random surfaces, Quantum Gravity
and Strings, Cargese, France (1990), Published in
the Proceedings.}
Next I will turn to a variety of questions in the second category,
mostly involving symmetries.  Recently there has been renewed
discussion of the plausibility of global symmetries, both
continuous and discrete.  In string theory,
it has been known for some time that there are no (unbroken)
continuous global symmetries.\refmark{\banksdixon}
String theory exhibits
intricate patterns of discrete symmetries, and it is often
conjectured that these symmetries are also gauge symmetries.
We will see, however, that this is not always the case.  String
perturbation theory often exhibits anomalous, global discrete
symmetries.  Such symmetries could be of great phenomenological
importance; they could, for example, lead to $m_u=0$.
(Other scenarios, with stringy features, which lead naturally to
$m_u=0$ have been considered in ref. \nirseiberg.)
These symmetries are closely related to certain gauged,
non-anomalous discrete symmetries.  These gauge symmetries are
spontaneously broken; the model-independent axion transforms
under them non-linearly.  As a result, they would seem to be
rather uninteresting.  In fact, however, we will see that using
these symmetries, one can hope to restrict the
form of any non-perturbative superpotential which might be
generated in string theory.  Indeed, we will see that
under these circumstances, one can argue
(subject to some plausible assumptions) that
``gluino condensation" is the largest supersymmetry-violating
effect at weak coupling.  These considerations may bear
on the possibility of a ``duality symmetry" involving the
dilaton, discussed by Joanne Cohn and John Schwarz at this
meeting,\refmark{\dilatonduality} as well as on the
general question of ``stringy
 non-perturbative effects."\refmark{
\shenker}

\REF\march{M. Kamionkowski and J. March-Russell, Phys. Lett.
{\bf 282B} (1992) 137;  R.~Holman {\it et al.},
Phys. Lett. {\bf 282B} (1992) 132;
S.M. Barr and D. Seckel, Bartol preprint BA-92-11.}
\REF\nelson{A. Nelson, Phys. Lett. {\bf 136B} (1984) 387.}
\REF\barr{S.M. Barr, Phys. Rev. Lett. {\bf 53} (1984) 329.}
Finally, we will comment on $CP$ and the strong $CP$ problem.
We will see, first, that $CP$ is a discrete gauge symmetry in
string theory (it can be thought of as a combination of a
general coordinate transformation and a non-Abelian gauge
transformation in the higher dimensional space).  As a result,
there can be no non-perturbative $CP$-violating parameters.
$CP$-violation is necessarily spontaneous, and
any $CP$-violation is in principle, calculable.  It is
probably premature to develop a detailed theory of the
origin of the $KM$ angle in string theory.  But it is of interest
to consider how the strong  $CP$ problem might
be resolved in string theory.  We have already
remarked that string theory may be a framework in which
to understand $m_u=0$.  We will comment on various aspects
of axions in string theory, particularly in light of recent
criticisms of the axion idea.\refmark{\march}
We will also see that, quite generally
in the framework of supersymmetric models (not just strings),
it is difficult to implement scenarios of the Nelson-Barr type,
in which $\theta$ is arranged to be zero at tree level.\refmark{\nelson,
\barr}
The problem
is that generically radiative corrections to $\theta$ are unacceptably
large, unless the soft breakings satisfy certain striking
constraints.\refmark{\dkl}

\chapter{Supersymmetry Breaking and the Problem of the
Dilaton}

\REF\gauginocond{M. Dine, R. Rohm, N. Seiberg
and E. Witten, Phys. Lett. {\bf 156B} (1985) 55;
J.P. Derendinger, L.E. Ibanez and H.P. Nilles, Phys. Lett.
{\bf 155B}(1985) 65}
It is appropriate to begin by reviewing the suggestions which
have been made for how supersymmetry breaking might
arise in string theory.  These revolve large around the
dynamics of ``gaugino condensation" in some hidden sector
gauge group.\refmark{\gauginocond}
The basic idea is very simple.  Suppose that
below the Planck scale the model contains a gauge group,
${\cal G}$, under which none of the matter fields transform.
The scale of the group is
$$\Lambda \sim e^{-{8 \pi^2 \over g^2 b_o}}.\eqn\hiddenscale$$
At this scale one expects that
gluinos condense, i.e.
$$<\lambda \lambda> \propto \Lambda^3 .\eqn\gluinocondensate$$
In the low energy theory, the dilaton superfield, $S$, couples
universally to the gauge fields.  In particular, the auxiliary
component of $S$, $F_S$, couples to $\lambda \lambda$,
so one finds a potential
$$V \sim \vert <\lambda \lambda> \vert^2.\eqn\potential$$
However, in string theory, the coupling, $g$, is itself determined
by the expectation value of the dilaton, $g \sim
e^{-D}$.  Thus $V$ is a potential for the dilaton.
This potential can be understood as arising from a non-perturbatively
generated superpotential of the form
$$W \sim  e^{-S/b_o}\eqn\nonperturbativew$$
where $S=e^{-D}+ia$, $a$ being the axion field.

\REF\dilatonproblem{M. Dine and N. Seiberg,
Phys. Lett. {\bf 162B}, 299 (1985).
and in {\it
Unified String Theories}, M. Green and D. Gross, Eds. (World Scientific,
1986).}
Unfortunately, however, while the appearance of this superpotential
constitutes dynamical supersymmetry breaking, at least at weak
coupling, this potential has no minimum except at zero coupling,
i.e. infinite value of the dilaton vev.\refmark{\dilatonproblem}
One can object that perhaps this problem is an artifact of our
assumptions.  Indeed, except in special cases (to be discussed
below) there is no argument that high energy, non-perturbative
string effects cannot be {\it larger} than any effects which
can be seen in the low energy theory.  For example, integrating
out massive string modes might lead to a superpotential
with a different dependence on the dilaton than that of eqn.
\nonperturbativew.  A little thought, however, makes clear
that the problem of the runaway dilaton is generic;
the potential will always go to zero
in weak coupling.  One can hope that there is a minimum at strong
coupling.  Such a prospect, however, is disturbing, not merely
because we have at present no idea how to treat strongly coupled
string theories, but also because it is precisely the features of
string theory at weak coupling which make it so attractive.

\REF\krasnikov{N.V. Krasnikov, Phys. Lett. {\bf 193B} (1987)
37; L. Dixon, in {\it Porceedings of the DPF Meeting},
Houston, 1990 : J.A. Casas, Z. Lalak, C. Munoz and G.G. Ross,
Nucl. Phys. {\bf B347} (1990) 243; T. Taylor, Phys. Lett.
{\bf B252} (1990) 59.}
There has been at least one interesting proposal to remedy
this problem.\refmark{\krasnikov}
Suppose one has several hidden sector groups,
each producing a gluino condensate.  Then, for some choices of
phase of the condensate, the non-perturbative superpotential
has the form
$$W= \alpha e^{-a S} -\beta e^{-b S}. \eqn\dualingcondensates$$
Now, because of the negative sign,
one can have a local minimum in the potential.  If $a \approx b$,
then it is possible that this minimum lies at large $S$, i.e.
weak coupling.
Note, however, that there is no parameter at our disposal
which can be made arbitrarily small.  Instead, one tries to
find compactifications which lead to numerically large values
of $S$, values which, from a field-theoretic perspective, one expects
to correspond to weak coupling. \foot{One can debate
whether, in the absence of a small parameter which can
formally be taken to zero,
there can be a sense in which weak coupling is valid.
To fully justify this procedure one must be able
to argue that even though the effects of the high energy
theory are not under control, one can still integrate
them out, obtaining an effective action of a known form,
and that the largest supersymmetry breaking effects
are those which are visible in the low energy theory.
In the case of anomalous discrete symmetries, discussed
below, this may not be necessary.}
In practice, examples have been constructed with rather small
effective coupling.  However, even though supersymmetry
turns out to be broken in some of these minima, the cosmological
constant is typically non-zero even in the leading approximation.

\chapter{The Minimal Supersymmetric Standard Model}

The MSSM, as noted in the introduction, has become the
paradigm for a supersymmetric model of nature.  In this
section, I would like to review the features of this model.
We will see that the basic assumptions of the model are
very strong, but that if supersymmetry breaking in string
theory takes one very particular form, string theory can
yield precisely such a theory.

The particle content of theMSSM is that of a
supersymmetrized ordinary standard model with two Higgs
doublets (i.e. one takes the gauge group to be $SU(3) \times
$SU(2)$ \times U(1)$,
introduces a chiral superfield for every
quark and lepton, and introduces two $SU(2)$ doublet
fields, $H_1$ and $H_2$).
Of course, in string models one typically obtains additional fields,
such as gauge singlet chiral fields and additional gauge interactions,
but such a spectrum might emerge from a string model.

\REF\masiero{F. Gabiani and A. Masiero, Nucl. Phys. {\bf B322}
(1989) 235}
To specify the model, it is also necessary to make some statements
about soft breakings.  There are a variety of experimental
constraints on these parameters.  Obviously, the masses of the
superpartners must be large enough that these particles
not have been seen.  Presumably, these masses should not be
arbitrarily large, since in that case it will be necessary to fine
tune parameters in order to obtain electroweak breaking.
Operationally, the most severe constraint of this type arises
from corrections to the $H_1$ mass containing a top squark;
these are proportional to $3m^2_{\tilde t}  g_t^2/(16 \pi^2)$
times a logarithmic factor.
Here $g_t$ is the top quark Yukawa coupling, and $m^2_{\tilde t}$
is the top squark mass.
One also requires significant degeneracy of squark masses to avoid flavor
changing neutral currents
(FCNC's).  Indeed, from the real part of the $K -\bar K$ mass
difference, one obtains a constraint (ref. \masiero)
$${\delta m^2 \over m_{susy}^2} < 10^{-2}-10^{-3}\eqn\kkbarlimit$$
where $\delta m^2$ represents a typical splitting and $m_{susy}$
a typical susy-breaking mass.

\REF\kanereview{A good overview of the positive features
of supersymmetry and supersymmetric unification, with
extensive references, is
provided by G. Kane, UM-TH-93-10.}
\REF\amaldi{P. Langacker and M. Luo,
Phys. Rev. {\bf D44} (1991) 817; U. Amaldi, W. de Boer and
H. Furstenau, Phys. Lett. {\bf B260} (1991) 447.}
Without specifying further details of the soft breakings, it should be
noted that this model has already scored some striking
successes.\refmark{\kanereview}
Perhaps most dramatic is the successful unification of
gauge (and some Yukawa) couplings which results in this
picture.\refmark{\amaldi}
Other good features include the presence of suitable
dark matter candidates.

Of course, the model as it stands possesses a large number of
free parameters, and a number of theoretical approaches have been
adopted to narrowing this parameter space.  The most common
assumption is that all of the squarks and sleptons are degenerate
at the unification scale,
and that the soft-breaking cubic couplings are simply proportional
to the superpotential.  One then evolves to low energies
using the renormalization group, trading one parameter for
$M_Z$.  This leaves four parameters (excluding $m_t$).
Further constraints which are often imposed include:
$m_b=m_{\tau}$ at the high scale; presence of a suitable dark
matter candidate, and absence of fine tuning.  Such programs
lead to suggestive values of the soft-breaking
parameters.\refmark{\kanereview}

\REF\il{L. Ibanez and D. Lust, Nucl. Phys. {\bf B382}
(1992) 305.}
These are strong assumptions, and they are not easy to justify within
the most popular framework for supersymmetry model building:
supergravity theories with supersymmetry broken in a hidden
sector.  In such models one has a set of ``hidden sector fields,"
$\Phi$ and ``visible sector fields, " $Q^I$(quarks,
leptons, etc.; here we are adopting the notation of ref.
\vadim).  The general theory of this type is
specified by three functions:  the Kahler potential, $K(\phi,\phi^*)$,
the superpotential, $W(\phi)$, and a function $f(\phi)$ which
describes the coupling of matter to gauge fields.  In this approach,
one assumes that some of the hidden sector auxiliary fields obtain
vacuum expectation values, $F_{\Phi} \sim m_{3/2} M_p$.
Taylor expanding the Kahler potential about the origin of the
visible sector fields,
$$K=K(\Phi^*,\Phi) +Z_{\bar IJ}(\Phi^*,\Phi)Q^{* \bar I} Q^J
+({1 \over 2} H_{IJ}Q^I Q^J + cc.) + ...\eqn\kahlerexpansion$$
The term $Z_{\bar I J}$ is the origin of the squark masses.
Squark
degeneracy means
$$Z_{\bar I J} \propto \delta_{\bar I
J}.\eqn\degeneracycondition$$
But there is no reason, in general, for this condition to hold;  there
is certainly no symmetry of the low energy theory which enforces it.
There is no reason one would expect this to hold generically in
string theory; indeed, Ibanez and Lust\refmark{\il} have
shown in particular orbifold examples that such a relation does not
hold.

While a number of suggestions have been made over the years to
understand degeneracy, Kaplunovsky and Louis have recently pointed out a
possible stringy solution.\refmark{\vadim}
These authors tried to study the question
of soft breakings in the context of string theory, without making
detailed assumptions about the origin of supersymmetry breaking.
They assumed only
\pointbegin
The potential has a stable minimum at reasonably weak coupling.
\point
$V \approx 0$ at the minimum
\point
SUSY breaking is primarily due to expectation values either for
the moduli or the dilaton, $<F_M>$ or $<F_S>$.

These assumptions are modest in the sense that they are
probably the
minimal assumptions required for any plausible phenomenology;
whether or not these conditions are actually achieved in string
theory is another matter.
Without further assumptions, one can make only modest
statements (see V. Kaplunovsky's talk at this meeting).
In particular, there is no explanation for squark degeneracy.
However, these authors noted that if one assumes
$<F_S> \gg <F_M>$, then, because of the universality of the dilaton
couplings, one {\it does} obtain a significant degree of squark
and slepton degeneracy, and one in fact obtains a highly predictive
scenario.  Examining the tree level lagrangian, one finds that
the gaugino masses, $m_g$, the squark and slepton masses,
$m_{\phi}^2$, and the $A$ parameter (the coefficient
of the cubic soft breaking terms in the potential) are given by
$$m_g= {\sqrt 3 \over 2} m_{3/2}~~~~m_{\phi}^2 = m_{3/2}^2
{}~~~~A=-\sqrt{3} m_{3/2}.\eqn\dilatonparameters$$

\REF\barbieri{R. Barbieri, J. Louis and M. Moretti, CERN-TH.6856/93}
This is precisely the structure of the soft-breaking parameters in
the minimal supersymmetric standard model!  Moreover, instead
of the four parameters listed above, there is now only one
(or two, depending one what one assumes about the origin of
the so-called $\mu$ parameter).\refmark{\vadim}
One expects that
these relations will be corrected by one loop effects of order
$\alpha_{GUT}/\pi$.  Whether or not this is good enough to
explain the suppression of FCNC's
will depend on the precise value of the one loop coefficients.
Naively, however, one does not expect these corrections to
be small enough, since at the large scale $\alpha_s /\pi$,
for example, is likely to be larger than $10^{-2}$, so
something more is likely to be required, particularly to
understand the smallness of the imaginary part of $K \bar K$.
Still, this looks tantalizingly close.

We do not know, of course, whether string theory dynamics
satisfy this condition.  Kaplunovsky and Louis, in fact, argue
that they do not for known susy breaking schemes.
However,
we know very little about susy breaking in string theory;
no known mechanism even satisfies the modest set
of assumptions listed above.
Given that the dilaton breaking scenario looks close to what
one wants,
and that only one assumption is required, it is
interesting to explore its consequences.  The required
renormalization group analysis has been performed in ref. \barbieri.
Assuming MSSM particle content, one finds that it is difficult
to implement this scenario given current experimental
constraints; significant fine tunings (at the part in $10^{-2}$
level) are required.  Non-minimal models are presumably
not so highly constrained (and of course not so predictive);
it will be of interest to explore their phenomenology.

\chapter{Flavor Symmetries:  An Alternative Solution to the
Degeneracy Problem}

An alternative approach to the problem of degeneracy,
about which string theory also offers some suggestive clues,
is to assume that there is some underlying flavor
symmetry.\refmark{\dkl,\nirseiberg}
One possibility is that there exists a
non-Abelian, gauged horizontal symmetry.
The  obvious problem with this proposal
is that whatever flavor symmetry there is must be badly
broken in order to explain the ordinary quark mass matrix.
The simplest model which one can use to examine this question
possesses an extra gauge symmetry $SU(2)_H$.
Under this symmetry we assume that the quarks transform as
doublets and singlets:
$Q^a,\bar u^a,\bar d^a$ and
singlets, $Q_s,  u_s, \bar d_s$
and similarly for the leptons.  The Higgs fields are assumed
to be $SU(2)_H$ singlets.

In addition, to break the horizontal
symmetry, we assume that we have some $SU(2)_H$ doublets
which are standard model singlets, $\Phi^a_{(i)}$.
{}From a stringy perspective, the presence of such particles
is quite plausible.  We might imagine that the $SU(2)_H$
symmetry corresponds to an enhanced gauge symmetry
at some particular point in the moduli space.  The fields
$\Phi^a_{(i)}$ then represent moduli.  Giving them expectation
values corresponds to moving away from the special point.
If this identification is correct, these fields have no potential,
and can easily obtain vev's of order $M_p$.  For what
follows, we will assume
$${<\Phi^a_{(i)}> \over M_p} \sim 10^{-1}.\eqn\modulivevs$$
Such an assumption is not unnatural.  Vev's of this order
might arise in the presence of Fayet-Iliopoulos terms generated
at one loop, for example.

\REF\cargese{G. 't Hooft, in {\it Recent Developments
in Gauge Theories}, G. 't Hooft et. all, Eds., Plenum
(New York) 1980.}
To proceed, we need to adopt a set of rules about the
sizes of various couplings.  We will enforce 't Hooft's
notion of naturalness:\refmark{\cargese}
couplings can be small only if the theory becomes
more symmetric in that limit.  Ultimately, we would
like to explain such small parameters through additional
symmetries (e.g. discrete symmetries of the type to be
discussed shortly).

Let us consider, then, the allowed couplings in the lagrangian.
The superpotential just below $M_p$ contains
dimension-four terms:
$$W_q= \lambda_1 \epsilon _{ab} Q_a \bar d_b H_1 + \lambda_2 \epsilon _{ab} Q_a
\bar u_b H_2 + \lambda_3 Q_s \bar d_s H_1
+ \lambda_4 Q_s \bar u_s H_2 .\eqn\wdimensionfour$$
These give rise to $SU(2)_H$ symmetric terms in the mass matrix.
Clearly we need to assume that $\lambda_1$ and $\lambda_2$
are small (this might be arranged by means of a discrete symmetry).
$SU(2)_H$-violating terms arise at the level of dimension five
and dimension six operators:
$$
{1 \over M_p}(\lambda_5^i \epsilon _{ab}\Phi_a^i  Q_b \bar d_s H_1
+ \lambda_6^i \epsilon _{ab}\Phi_a^i  Q_s \bar d_b H_1)
+{1 \over M_p^2}(\lambda_7^{ij} \epsilon _{ab}\epsilon _{cd}\Phi_a^i \Phi_c^j
Q_b \bar d_d H_1 +.....).
\eqn\dimfiveandsix$$
Note that the charmed-quark mass must arise from these
operators, and is thus of order $(\Phi/M_p)^2$,
so $\Phi/M_p$ can't be much smaller than $0.1$.

The breaking of the squark degeneracy can also be
understood in terms of the effective action at scales
slightly below $M_p$.  This lagrangian contains
dimension-four, soft-breaking terms which
give $SU(2)_H$-symmetric contributions to the squark
mass matrices:
$$V_{soft}= m_1^2 \vert Q_a \vert^2 + m_2^2 \vert Q_s \vert^2+
m_3^2 \vert \bar u_a \vert^2 + m_4^2 \vert \bar u_s \vert^2 + ...$$
$$+A_1 \lambda_1 Q \bar d  H_1+A_2 \lambda_2 Q \bar u H_1 + .... + h.c.
\eqn\softterms$$
Here, $m_i$ and $A_i$ are of order $m_{susy}$.
Breaking of the symmetry
will arise through terms of the type
$$\delta V^2_{soft}= {m_{susy}^2 \over M_p}(\gamma _1
\Phi _1 Q Q_s^* + ...)
+ {m_{susy}^2 \over M_p^2}(\gamma^{\prime}_1
\Phi_1 Q \Phi_2 Q^* + ...) \eqn\nonrensoftbil$$
and
$$\delta V^3_{soft} =  {m_{susy}\over M_p}
\lambda_5^1 Q \bar d_s  H_1 (\eta_1 \Phi_1
+  \eta _2 \Phi_2+ \eta_3 \Phi_2^*)$$
$$ +{m_{susy}\over M_p^2} \lambda_7^{11} Q \bar d H_1
(\eta '_1 \Phi_1 \Phi_1 +\eta '_2
 \Phi_1 \Phi_2 + \eta_3^{\prime} \Phi_1 \Phi_2^*)+\dots
 \eqn\nonrenormsofttril$$
We have omitted $SU(2)_H$ indices on $Q$, $\bar u$, $\bar d$ but terms
with all possible contractions should be understood.  Here $\gamma$,
$\gamma^{\prime}$, $\eta$ and $\eta^{\prime}$ are dimensionless
numbers. By 't Hooft's naturalness criterion,\refmark{\cargese}
many of these
couplings should not be much less than one; the theory does not become
any more symmetric if these quantities vanish.
As a result, the generic symmetry-violating terms in the first two
generations are of order $(\Phi/M_p)^2 \sim 10^{-2}$.
Some of these couplings, however,
can  (and should!) naturally be small.  So there is no difficulty with
suppressing FCNC's.

This framework is predictive.  For example, it suggests that there
should be a high degree of degeneracy only in the first two
squark and slepton generations.  The small value of the neutron
electric dipole moment is also readily accomodated here.
Recently, Seiberg and Pouliot have considered models
in which non-Abelian symmetries play a role both in producing
squark degeneracy and in explaining the features of the quark mass
matrix.  Such models are clearly of great interest.\refmark{\seibergrecent}

An alternative approach to the problem
of fcnc's in supersymmetry has recently been explored in ref.
\nirseibergb.  Here there is no degeneracy at all, but rather a tight
alignment between the quark and squark mass matrices.
This is done in the context of a more complete theory of fermion
masses with many stringy features.

 \chapter{Symmetries and String Theory}

\REF\gsw{M. Green, J. Schwarz and E. Witten, {\it Superstring Theory},
Cambridge University Press, New York, 1986.}
String theory has much to say about the question of symmetries
which might plausibly appear in a low energy effective field theory:
\pointbegin
Global vs. local continuous symmetries:
It has long been argued that it doesn't make much sense to
impose global symmetries in field theory.  In string theory,
this prejudice takes on the status of a theorem:  one can
show that string theory posseses no global continuous
symmetries.\refmark{\banksdixon}
\point
Discrete symmetries:  String models often exhibit a rich structure
of discretesymmetries.\refmark{\gsw}
These can often be thought of as ``gauge symmetries," e.g. coordinate
or gauge transformations in some higher dimensional theory.
\point
Matter multiplets in string theory do not typically have the structure
of conventional grand unified theories.\refmark{\gsw}
In particular, the
transformation properties of fields under discrete symmetries do
not correspond to those of GUT multiplets.

\REF\krauss{L. Krauss and F. Wilczek, Phys. Rev. Lett. {\bf
62}, 1221 (1989).}
\REF\ir{L. Ibanez and G. Ross, Phys. Lett. {\bf 260B} (1991) 291;
Nucl. Phys. {\bf B368} (1992) 3.}
\REF\preskill{J. Preskill, Sandip Trivedi, F. Wilczek and M. Wise,
Nucl. Phys. {\bf B363} (1991) 207.}
\REF\banksdine{T. Banks and M. Dine, Phys. Rev. {\bf D45} (1992)
424.}
Discrete symmetries are of great phenomenological importance
in supersymmetric model building:  they are needed to forbid
proton decay, and   perhaps other rare processes
($\mu \rightarrow e \gamma$, etc.).  They appear in many
other possible extensions of the standard model, as well.
It has been argued that any discrete symmetries appearing in
an effective lagrangian should be gauge
symmetries.\refmark{\krauss}  Otherwise,
they are likely to be spoiled by gravitational effects.
It has also been stressed that discrete symmetries may
be anomalous;\refmark{\ir,\preskill} this suggests
anomaly constraints on discrete symmetries.  In ref. \ir,
these constraints were enumerated assuming that the discrete
symmetries were embedded in a broken continuous gauge group,
and that
the charges of the heavy states were integer multiples of those
of the light states.  This lead to a quite strong set of constraints.
In general, however, only a weaker set of constraints hold;
these can be understood in terms of instantons in the low
energy theory.\refmark{\ir,\banksdine}

In ref. \banksdine, it was noted
that in string theory there {\it are} often anomalous discrete
symmetries.
However, one can always cancel these anomalies by assigning
to the (model-independent) axion a non-linear transformation law
under the symmetry,
$$a \rightarrow a + 2 \pi \delta.\eqn\axiontransformation$$
($\delta$ would be a multiple of $1/N$ for a $Z_N$ symmetry).
(This possibility had been suggested in ref. \ir.)
Such a transformation law means, of course, that the discrete
symmetry is spontaneously broken (at a scale of order $M_p$).
However, this observation has another consequence:
in perturbation theory, in such cases, there is an unbroken,
{\it global}, anomalous discrete symmetry.  After all, perturbation
theory exhibits an unbroken (spontaneously
or explicitly) discrete symmetry .

\REF\gsw{M. Green, J. Schwarz and E. Witten, {\it Superstring
Theory}, Cambridge University Press, Cambridge (1987).}
It is perhaps helpful to give a simple example of this phenomenon
(which was discussed in ref. \banksdine).
Such an example
is provided by the compactification of the heterotic string
on the Calabi-Yau manifold associated with the quintic hypersurface
in $CP^4$, discussed, for example, at some length in the
textbook of Green, Schwarz and Witten.\refmark{\gsw}
At certain points in the moduli space, this model possesses
a freely-acting
$Z_5 \times Z_5$ symmetry.  In the textbook treatment,
one mods out by this symmetry, including Wilson lines,
to obtain a model with a low number of generations and
a reasonable gauge group.  For our purposes (despite our title)
we are not concerned if the number of generations happens
to be large.  We can mod out by one of the $Z_5$'s, corresponding
to rotating the coordinates, $Z_a$, of $CP^4 $, by phases:
$$Z_a\rightarrow \alpha^a Z_a$$
where $\alpha=e^{2 \pi i /5}$.  This is still freely acting;
this means that we don't have to worry about
the appearance of massless particles in twisted sectors.

The main virtue of this choice is that it leaves over a set
of $R$-symmetries.  For definiteness, consider the
symmetry under which $Z_1 \rightarrow \alpha Z_1$.
Under this symmetry, the gluinos transform by a phase
$\alpha^{-1/2}$.  Now we can
include a Wilson line without breaking this symmetry.
For example, we can include a Wilson line in the ``second"
$E_8$ (the one which is unbroken in the absence of the
Wilson line), described by:
$$a= {1 \over 5}\left ({1,1,2,0,0,0,0,0}\right ).\eqn\wilsoncy$$
(I am using the notation which is standard in the orbifold
context).  By itself, this choice is not modular invariant, but
this is easily repaired by including a Wilson line in the first
$E_8$ as well.  In the second $E_8$, there are
two unbroken gauge groups.  It is easy to determine
the effects of instantons by simply examining $SU(2)$
subgroups of these.  One finds that instantons
of the first group have four gluino zero modes, while
instantons of the second have $24$.  Thus assigning
to the axion a transformation law
$$a \rightarrow a + {12 \pi \over 5}\eqn\funnydelta$$
(the reason for writing $12$ rather than $2$ will
become clear shortly)
cancels both anomalies.

This observation is potentially of great importance for model
building.  It suggests that it is reasonable to impose anomalous,
global discrete symmetries.  For example, one might
impose a symmetry which forces $m_u=0$ in order to solve
the strong $CP$ problem.  (Recall that the discrete symmetries
of string theory don't have the structure of those of conventional GUT's.)

\REF\dlm{M. Dine, R. Leigh and D. MacIntire, in preparation}
These anomalous discrete symmetries are interesting from
at least one other viewpoint.  They should permit one to make
{\it exact non-perturbative} statements about string
dynamics.\refmark{\dlm}
The reason is simple.  The axion and dilaton together make
up the complex scalar in a supermultiplet, usually denoted
$S$,
$$S= {1 \over g^2} + ia.\eqn\dilatonsuper$$
Given the axion transformation law of eqn. \axiontransformation,
$W(S)$ must take the form
$$W(S)= \sum_n C_n(M)e^{-n/\delta}.$$
This is precisely the behavior one encounters in the gluino
condensation scenario.  In these cases, one can argue
that there can be no stringy non-perturbative effects
stronger than the effects observed in the low energy field theory!
The main subtlety in this argument lies in the determination
of $\delta$.  $\delta$ cannot be determined unambiguously from
the anomaly considerations described above; typically one can
add a constant of the form $2 \pi r$, where
$r$ is an integer.  For example, had we chosen
$2$ rather than $12$ in eqn. \funnydelta,
the superpotential of gluino condensation would
not have been invariant.  This issue will be
discussed further in ref. \dlm.

Can one generalize this?  The discussion here is close in spirit
to discussions of ``dilaton duality" which have appeared
recently in the literature.\refmark{\dilatonduality}  I suspect that
some version of this duality is indeed correct; however
it is difficult to establish it by the sort of arguments which
have been employed here.  (See the talks by Joanne Cohn
and John Schwarz at this meeting.)  These issues will also
be discussed in ref. \dlm.

\chapter{String Theory and the $\theta$ Puzzle}

\REF\muzero{H. Georgi and I. McArthur, Harvard
University Report No. HUTP-81/A011 (unpublished);
D.B. Kaplan and A.V. Manohar, {\it Phys. Rev. Lett.} {\bf
56}, 2004 (1986); K. Choi, C.W. Kim and W.K. Sze, {\it Phys. Rev. Lett.}
{\bf 61}, 794 (1988); J. Donoghue and D. Wyler, {\it Phys. Rev.} {\bf D45}
(1992) 892; K. Choi, Nucl. Phys. {\bf B383} (1992) 58.}
There are three known solutions to the strong CP problem.
\pointbegin
$m_u=0$.  This solution may be consistent with current
algebra.\refmark{\muzero}
As we have noted in the previous section, this is a prediction
which could quite plausibly emerge from string theory.
\point
Peccei-Quinn symmetries and axions:  here one needs a
global symmetry to a high degree of approximation in perturbation
theory.
\point
Spontaneous CP violation:  in theories with an exact $CP$
symmetry, the bare $\theta$ is zero.   The problem is then
to arrange that $\theta$ be small at tree level after
symmetry breaking, and that radiative corrections be sufficiently
small.\refmark{\nelson,\barr}

\REF\strominger{A. Strominger and E. Witten, Comm. Math. Phys.
{\bf 101} (1985) 341.}
String theory has interesting statements to make about all three
possibilities.  To consider these, we should first understand
the status of $CP$ as a symmetry of string theory.
Strominger and Witten were probably the first to comment
on $CP$ in string theory.\refmark{\strominger}  These
authors noted that certain Calabi-Yau compactifications
possess unbroken CP at points in their moduli spaces.  $CP$,
then, is a symmetry of the classical theory
which can be spontaneously broken
by vev's for moduli.  It is not hard to see that in these
cases, $CP$ is a good symmetry of perturbation theory,
since it corresponds to a two-dimensional symmetry which
commutes with the $BRS$ operator.

\REF\ckn{K. Choi, D. Kaplan and A.E. Nelson,
Nucl. Phys. {\bf B391} (1992) 515.}
\REF\dlmcp{M. Dine, R. Leigh and D. Macintire,
{ Phys. Rev. Lett.} {\bf 69} (1992)
2030.}
What about non-perturbatively?  For example, there has been
speculation on the possibile existance of non-perturbative
parameters in string theory.  Could some of these exist
and violate $CP$ ($\theta$, after all, is the quintessential
non-perturbative parameter)?  While we don't understand
much about non-perturbative string theory, the answer
to this question is a definite ``no."  $CP$ turns out
to be a gauge symmetry in string theory, a combination
of general coordinate and gauge transformations in the
higher dimensional space.\refmark{\ckn,\dlmcp}

To see this, consider the $O(32)$ or $E_8 \times E_8$
heterotic strings in ten dimensions.  In ten dimensions,
$P$ is not a symmetry of either theory.  What one would
like to call $C$ has the effect of changing the signs of
all of the lattice momenta, while acting trivially on
spinors (as a consequence of GSO).  This change of signs
is a symmetry of the lattice; in fact, it is a gauge transformation.
In $O(32)$, this transformation acts on a
$32$ component vector by the matrix:
$$\Lambda = \left ( \matrix{i \sigma_2 & 0 &   \dots\cr  0 &
i \sigma_2  & \dots \cr \dots & \dots &\dots }
\right ).\eqn\cprotation$$
A similar transformation works in $E_8$.

Now compactify to four dimensions on an ordinary torus.  In
this theory, it is well-known that $P$ is a symmetry.  Where
did this come from?  Denote the compactified coordinates
by $y^a$, $a= 1 \dots 6$, and the uncompactified coordinates
by $x^{\mu}$.  The transformation
$$x^i \rightarrow -x^i ~~~y^2 \rightarrow -y^2~~~
y^4 \rightarrow -y^4~~~y^6 \rightarrow -y^6\eqn\fourdp$$
is a good space-time symmetry (it commutes with GSO).
{}From the perspective of ten-dimensions, it is part of the
proper Lorentz group.   Thus for these compactifications,
$CP$ is a product of an ordinary transformation and a general
coordinate transformation (with a little work, you can check
that this transformation has the expected action on spinors).
Other compactifications, such as Calabi-Yau compactifications,
typically preserve this product of symmetries at some points
in the moduli space.

$CP$ is thus special.  It can't be broken by no-perturbative effects
(instantons, wormholes, etc.); non-perturbative, $CP$-violating
parameters cannot arise in string theory, and in particular
there can be no bare $\theta$ (e.g. in the $E_8 \times
E_8$ theory).  Thus all $CP$-violating effects in string theory
are, in principle, calculable.

An obvious question is whether one can use this to implement
the spontaneous $CP$-violation solution to the strong $CP$-problem.
A few observations are important.  First, to accomplish
this, one wants $CP$ broken at a scale well below
$M_p$.\refmark{\ckn}  If $\Lambda$ is the scale
of $CP$ violation, even if one suppresses low
dimension operators contributing to $\theta$, there will
inevitably be operators of high dimension, whose contribution
will be of order $(\Lambda/M_p)^n$.  Thus one probably wants
to break $CP$ at a scale not larger than $10^{10}$ GeV, or so
(assuming that the leading contributions are due to dimension
five operators).
Otherwise, the effective $\theta$
is almost certainly of order one.\refmark{\ckn}

\REF\frampton{P.H. Frampton and T.W. Kephart,
Phys. Rev. Lett. {\bf 65} (1990) 1549.}
\REF\dkmns{M. Dine, V. Kaplunovsky, M. Mangano, C. Nappi and
N. Seiberg, Nucl. Phys. {\bf B259} (1985) 549.}
$E_6$ models, such as those which arise
in $(2,2)$ compactifications of string theory, provide a
rather natural setting for the Nelson-Barr
mechanism.\refmark{\frampton}  Under
$O(10) \times U(1)$, the $27$ of $E_6$ decomposes as
$$27=16_{-1/2} + 10_1 + 1_{-2} .\eqn\twentysevendecomp$$
Here the $16$ contains an ordinary generation of quarks
and leptons and an additional right-handed
neutrino, ${\cal N}$, while the ten contains a vectorlike (with
respect to the standard model group) set of doublets,
$H_1$ and $H_2$, and
a vectorlike set of charge $-1/3$ quarks, $q$ and $\bar q$.
We will denote the $O(10)$ singlet by $S$.  The $SU(3)\times
SU(2) \times U(1)$ singlet fields, ${\cal N}_i$ and $S_i$
can naturally obtain vev's of order\refmark{\dkmns}
$$M_{INT}=\sqrt{m_W M_p} \approx 10^{11} GeV.\eqn\intscale$$
Moreover, there is typically a range of soft-breaking
parameters for which the $<S_i>$ are real, while the
$<{\cal N}_i>$ are complex.  The allowed couplings in
the superpotential in these models are of the form
$$W = {\cal N} q \bar d + S q \bar q
+ H_1 Q \bar u + H_2 Q \bar d. \eqn\allowedw$$
The fermion mass matrix then has the Nelson-Barr structure:
$$m_F= \left ( \matrix{ \Gamma H_2 & \gamma {\cal N} \cr
0 & \mu} \right ). \eqn\barrnelsonmatrix$$
Automatically, $arg~det~m_F=0$.

\REF\barrmasiero{S.M. Barr and A. Masiero, {\it Phys. Rev.}
{\bf D38} (1988) 366;S. Barr and G. Segre, preprint BA-93-01 (1993)..}

At tree level, there are many couplings which must be
suppressed for all of this to work.  For example, it is
necessary to avoid phases in the Higgs potential, and
to suppress couplings which would lead to phases in the
$S$ vev's.  This can be accomplished by discrete symmetries.
In loops, however, more serious problems arise which
cannot be dealt with so easily.  Early studies of loop
corrections to $\theta$ showed that many contributions can
be suppresed.\refmark{\barrmasiero} However, these analyses
assumed that squarks are precisely degenerate.
If one relaxes the assumption of exact degeneracy and examines the
various corrections
one finds that the smallness of $\theta$ sets stringent
requirements.  In particular, diagrams correcting
the $d$ quark mass lead to a requirement that
${\delta m^2 \over m_{susy}^2} < 10^{-5}$;
diagrams contributing to the gluino mass require that
certain phases be aligned to one part in $10^{7}$.  Neither
of the schemes we have discussed above (flavor symmetries or
dilaton driven supersymmetry)
seem likely to produce anything like this degree of degeneracy.
Perhaps these constraints can be satisfied as a consequence of symmetries.  But
at the least a much more elaborate symmetry
 structure is required to implement the Nelson-Barr scheme
in the framework of supersymmetry than has been
considered to date.

\chapter{Axions in String Theory and Elsewhere}

The underlying idea to the axion solution to the strong CP
problem is to postulate a ``Peccei-Quinn symmetry"
under which the axion transforms as
$$a \rightarrow a + f_a \delta.\eqn\axiontransformation$$
This solution, however, suffers from serious problems,
which, like the Barr-Nelson solution discussed
above, call its plausibility into question.\refmark{\march}

The basic problem is simple, and should be ``obvious" to string
theorists.  In string theory, and presumably in any fundamental
theory, there are no global, continuous symmetries.  Approximate
global symmetries, if they exist,
might arise accidentally, as an accidental consequence
of the structure of low dimension terms in the low energy
effective lagrangian (like $B$ and $L$ in the standard model).

To see the difficulty, recall that the usual axion potential
has the form
$$V=-N m_{\pi}^2 f_{\pi}^2 cos(a/f_a). \eqn\axionpotential$$
Suppose that ${\cal O}_n$ is the lowest dimension operator
which breaks the Peccei-Quinn symmetry;  its dimension is
$n+4$.  Then the symmetry-breaking lagrangian has the
form
$${\cal L}_{SB}= {\gamma \over M_p^n} {\cal O}_n.\eqn\lsb$$
This gives rise to an axion potential, on dimensional grounds,
of the form
$$\delta V = {\gamma f_a^{n+1} \over M_p^n} a(x)
.\eqn\detav$$
If $f_a \sim 10^{11} GeV$, then requiring $\theta=a/f_a < 10^{-9}$
gives $n>7$, i.e. it is necessary to suppress operators up to
dimension $11$ in order to obtain a sufficiently good symmetry!

\REF\wittenaxion{E. Witten, Phys. Lett. {\bf B149} (1984) 359.}
But string theory always has an axion in perturbation
theory.\refmark{\wittenaxion}  This axion arises from
the antisymmetric tensor field, $B_{\mu \nu}$, which
in four dimensions is equivalent to a scalar.  Moreover,
the decay constant of the axion is of order $M_p$.  From
the perspective of the argument given above, this is highly
suprising.  It can be understood in a variety of ways.
First, the couplings of the antisymmetric tensor are governed
by a gauge principle.  This insures that in perturbation theory,
$B_{\mu \nu}$ enters the effective lagrangian
only through the gauge-invariant field strength
$H_{\mu \nu \rho}$.\foot{One might naturally ask how
the axion potential due to non-Abelian gauge interactions
is consistent with this symmetry.  In fact, it is.  Moreover,
it is not hard to understand why only non-perturbative
effects give rise to this mass.  I thank Renata Kallosh,
Lenny Susskind for discussions of this and related
matters, which will appear elsewhere.}
Alternatively, one can understand this result directly
in terms of the string perturbation expansion.
At $k=0$, the vertex operator for the axion
is a total derivative.

While the existence of this axion is rather suprising
from a field theoretic perspective, it is not at all
clear that this field can solve the strong $CP$ problem.
First, its decay constant is of order $M_p$, which
poses cosmological difficulties.  Second, if there
is any sort of hidden sector gauge group, it is likely
to give mass to this axion.

\REF\lazarides{G. Lazarides, C. Panagiotakopoulos and Q. Shafi,
Phys. Rev. Lett. {\bf 56} (1986) 432.}
\REF\casas{J. Casas and G. Ross, Phys. Lett. {\bf 192B} (1987) 119.}
\REF\lou{M. Dine, SCIPP-92-27.}
Thus, if the string theory solves the strong $CP$ problem
by the Peccei-Quinn mechanism, another axion must arise by
accident as a consequence of a continuous or a
{\it discrete} gauge symmetry.  This latter possibility was
already considered some time ago in refs. \lazarides\ and
\casas.  These authors noted, as we have mentioned earlier,
that it is quite natural to obtain a scale $f_a \sim 10^{11} GeV$
in string theory.  For example, the field $S$ of eqn.
\twentysevendecomp\ can appear in the superpotential
only in a restricted way.  Because there is no mass for this
field, and because of its gauge quantum numbers, the
leading term which can give rise to a potential for the
field $S$ and the corresponding field $\bar S$ which
might arise from the $\overline{27}$, is
$$W= {1 \over M_p^2} S^2 \bar S^2.\eqn\intscalew$$
If the soft breaking terms include terms of the form
$$V_{soft} = -m_{3/2}^2 \vert S\vert^2 + ...\eqn\softintscale$$
then $S^2 \sim m_{3/2} M_p$.
In order that there be an approximate symmetry, one needs
to forbid many operators in $W$ and $V_{soft}$.
This is possible with rather simple discrete
symmetries.\refmark{\lazarides,\casas,\lou}

\chapter{Summary}

We have touched here on a large number of topics.
There are a few points which I hope you will take away
from this talk:
\pointbegin
Kaplunovsky and Louis have exhibited one is perhaps the first
example of something which might cautiously be referred
to as a string prediction.  In particular, string theory
offers a simple option for explaining squark and slepton
degeneracy; this framework makes a series of strong predictions.
\point
String theory offers several lessons for model building.
\item{a.}  Discrete symmetries in string theory
can be global and anomalous (relevant
for $m_u=0$?); discrete symmetries in string theory
don't look like those from conventional GUT's.
\item{b.}  $CP$ is an exact, gauge symmetry of string
theory.  But even though $CP$ violation is spontaneous
and the ``bare" $\theta$ vanishes, it is difficult to
solve the strong $CP$ problem without axions or $m_u=0$.
\point
String theory suggests alternative approaches to the problem of
squark degeneracy, based on both continuous and
discrete flavor symmetries.

\bigskip
\centerline{\bf Acknowledgements}
I would like to thank A. Kagan, R. Leigh, D. MacIntire,
and N. Seiberg for many helpful conversations about these subjects.

\refout
\end